# UPSTREAM TRAFFIC CAPACITY OF A WDM EPON UNDER ONLINE GATE-DRIVEN SCHEDULING

NELSON ANTUNES, CHRISTINE FRICKER, PHILIPPE ROBERT, AND JAMES ROBERTS

ABSTRACT. Passive optical networks are increasingly used for access to the Internet and it is important to understand the performance of future long-reach, multi-channel variants. In this paper we discuss requirements on the dynamic bandwidth allocation (DBA) algorithm used to manage the upstream resource in a WDM EPON and propose a simple novel DBA algorithm that is considerably more efficient than classical approaches. We demonstrate that the algorithm emulates a multi-server polling system and derive capacity formulas that are valid for general traffic processes. We evaluate delay performance by simulation demonstrating the superiority of the proposed scheduler. The proposed scheduler offers considerable flexibility and is particularly efficient in long-reach access networks where propagation times are high.

## 1. INTRODUCTION

The development of dynamic optical switching is widely recognized as an essential requirement to meet the anticipated growth in Internet traffic. Already, passive optical networks (PONs) are in use in many countries to provide high speed access and penetration is expected to increase rapidly in coming years. In this paper we consider the performance of the IEEE standardized EPON and its likely developments where fiber capacity is multiplied by the use of wavelength division multiplexing (WDM) and reach is extended by the use of optical amplifiers.

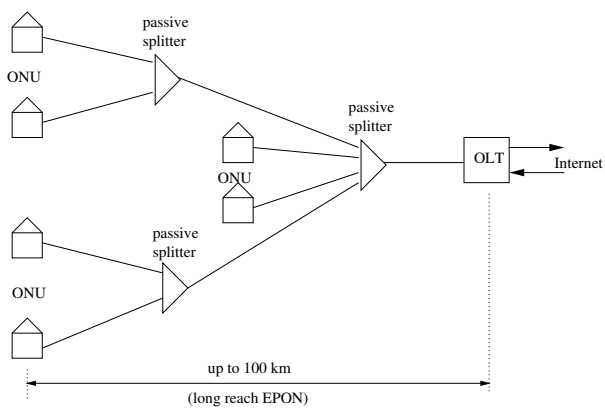

FIGURE 1. EPON components: ONUs, passive splitters and OLT

2         N. ANTUNES, C. FRICKER, PH. ROBERT, AND J. ROBERTSFigure 1 illustrates the main components of the EPON. Optical network units (ONUs) are situated in or close to user premises. An ONU might be dedicated to an individual domestic or business user or, when fiber is terminated at the curb or at the building, it could concentrate the traffic of several users. The ONUs of the EPON are controlled from an optical line termination (OLT) equipment that realizes the interface with the electronic packet switched network and manages downstream and upstream traffic. Passive splitters are used to broadcast downstream optical signals to the ONUs and to merge upstream signals destined to the OLT. There may be more than one level of splitters, notably in proposed long-reach PONs where the distance between central office and user can attain 100 km. The number of ONUs per EPON is currently limited to 64 for an EPON of upstream and downstream rate of 1 Gb/s. Future WDM EPONs will clearly have greater capacity.

Each ONU receives all downstream traffic on the wavelengths to which it has access and extracts its own packets based on the destination address. Upstream traffic management is more complicated. To avoid collisions, the ONUs must coordinate their upstream transmissions to the OLT. In EPON this coordination is performed by means of a dynamic bandwidth allocation (DBA) algorithm. The EPON implements a protocol allowing the ONUs to communicate their current requirements for upstream transmission time and the OLT to assign the necessary slots. The DBA thus emulates a kind of polling system. There are, however, a number of particularities that prevent us from directly applying known results on the performance of polling systems.

The physical separation of user queues (in the ONUs) and their controller (in the OLT) imposes a form of gated service with a variable delay between request and grant. This delay depends on the propagation time and is particularly significant for a long-reach EPON where the distance between OLT and ONU can attain 100 km. This leads to one-way propagation times of up to 500 $\mu$s which is significant compared to the transmission time of an Ethernet packet of only 12 $\mu$s at 1 Gb/s. It is necessary to compensate for the different propagation times of different ONUs. To ensure the polling cycle remains low even when some ONUs have high load, the DBA must impose an upper limit on the size of each grant thus realizing a kind of limited gated polling.

DBA algorithms may be distinguished as offline or online depending on whether or not the OLT waits to receive all requests before deciding on allocations and dispatching the grants. Offline DBAs allow tighter control and can implement sophisticated bandwidth sharing policies. It can, however, lead to excessive delays when propagation times are large since the schedule can only be computed once in every round trip time for the most distant ONU. Online DBAs are more efficient since the grant can be computed and sent as soon as the request is received. Most online algorithms described in the literature generate a grant explicitly for every request received. This policy can also lead to wasted capacity when propagation times are long since the service cycle is still at least as long as the longest round trip time. An alternative, that we explore in this paper, is for the OLT to issue grants spontaneously based on its current knowledge of ONU requirements. This is acquired from requests sent with upstream data sent in response to preceding grants. Polling is thus driven by grants rather than requests. After the names of the grant and request messages in EPON, we refer to these alternatives as GATE-driven and REPORT-driven scheduling, respectively. GATE-driven scheduling is simple to implement, compatible with standards and, to our knowledge, novel.

In the paper we describe GATE-driven scheduling and derive the traffic capacity it delivers for a classical single wavelength EPON and for a WDM EPON where ONUs access all wavelengths with full availability. Traffic capacity is defined as the load beyond which the underlying polling system would be unstable. We can determine this analytically under general traffic assumptions by adapting known results for multi-server polling systems. We evaluate delay performance by simulation and demonstrate that the GATE-driven algorithm clearly outperforms the classical REPORT-driven online scheduler. However, with the high transmission speed of optical fiber, delay performance is usually excellent until the network stability limit is attained. Traffic capacity is therefore the essential performance parameter for the EPON. Exact capacity results are more difficult to derive for WDM networks where ONUs have limited availability to the range of wavelengths. We propose and evaluate some approximations.

Related work. There is an abundant literature on EPON and the numerous DBA algorithms that have been proposed. Two recent surveys, by McGarry el al [18] and by Zheng and Mouftah [24], present a comprehensive review. Skubic et al. compare DBAs used in EPON with alternatives proposed for GPON, the alternative ITU standard. A significant early development was the inter-leaved polling algorithm IPACT proposed by Kramer et al. [11]. Most subsequently proposed DBAs seek to realize more complex sharing policies that, as discussed in the next section, we tend to consider less important than efficiently using the medium with low packet level latency. The algorithm we propose is closer to the original IPACT with limited gated service discipline.

DBA for the WDM EPON, sometimes called dynamic wavelength and bandwidth allocation (DWBA), has received much less attention. An early proposal by Kwong et al. is a WDM extension of IPACT using REPORT-driven scheduling [12].McGarry and various co-authors have notably compared offline and online DWBA algorithms [17, 16, 15] showing the superiority of the latter in terms of both capacity and packet delay. A quite complex DWBA algorithm is proposed by Dhaini et al. for a hybrid TDM/WDM EPON [5]. This is further developed and applied by Meng et al. in the context of a proposed joint access and metropolitan optical network architecture called STARGATE [19] Particularly relevant to the present work is a DBA proposed for a long reach GPON by Song et al. [21]. The authors recognize the resource waste caused by classical REPORT-driven scheduling and suggest the ONUs should initiate multiple interleaved cycles or threads. We believe our GATE-driven proposal is more efficient and offers greater flexibility.

Most performance evaluations of DBA algorithms rely on simulation. A few authors have developed analytical models generally using techniques developed for polling systems. Park et al. derived closed form formulas for the mean packet delay in a symmetric PON with identical ONUs and negligible propagation times under the assumption of Poisson traffic and gated service [20]. The two-stage polling system identified in that paper has been further analyzed by van der Mei and Resing [22, 23] when ONUs have heterogeneous load. A paper by Aurzada et al. demonstrates that classical polling models do not apply when propagation times are significant and the EPON scheduler is REPORT-driven [1]. This is because "switchover times" as the OLT successively sends grants to the ONUs are not then independent of the ONU queue sizes. This difficulty is removed with GATE-driven scheduling, as explained later. Moreover, since we focus on traffic capacity and not delay, we can make use of known results on the stability of multi-server polling systems that are applicable under quite general traffic assumptions.

The rest of the paper is organized as follows. In the next section we highlight some essential features of the EPON architecture. We then proceed in Section 3 to describe the GATE-driven scheme and to prove its claimed performance properties in the case of the classical single wavelength EPON. Section 4 is devoted to the WDM EPON. We derive the traffic capacity when the ONUs can use all wavelengths and discuss corresponding results when they have limited availability. The penultimate section is devoted to the presentation of simulation results where we numerically compare the delay performance of alternative DBA configurations.

## 2. WDM EPON ARCHITECTURE

We outline some basic architectural features deriving from technology and EPON standards before briefly discussing QoS.

2.1. **Technology.** The potential for WDM PON technology to realize a cost effective, high capacity access network is only beginning to be seriously explored. In particular, the issue of the number of wavelengths that can be economically maintained is hardly discussed. Although dense WDM (DWDM) is capable of creating hundreds of channels on a single fibre, we follow most works in assuming PONs will exploit a relatively small number of wavelengths.

An important issue is how the channels are shared. In several proposals, each ONU in a WDM PON is assigned a fixed pair of upstream and downstream wavelengths [9, 2]. With just one upstream wavelength per ONU the network is equivalent in terms of traffic capacity to a set of independent TDM PONs though some proposed architectures facilitate load balancing [8].

Maximum traffic capacity is obtained when all ONUs have full access to all wavelengths. However, this requires each ONU to be equipped with a costly array of fixed wavelength transceivers and network evolution to meet increasing demand is hard [13]. Nevertheless, this architecture constitutes a useful benchmark for measuring the performance of more cost effective alternatives.

Tunable transmitters exist that allow a source to switch wavelengths within a few nanoseconds. An ONU equipped with a tunable transmitter could thus access all available channels but use only one at any given time. A similar architecture results from the use of a reflective semi-conductor optical amplifier (RSOA) in the ONU [14]. The OLT transmits downstream a carrier burst on a chosen wavelength that is modulated by the ONU before being reflected back upstream to the OLT. The ONU is then particularly simple and inexpensive.

More generally, as advocated in [14], to allow progressive buildout and maximize flexibility, future WDM PONs should support ONUs with varied technologies and capabilities. Some ONUs might have one or several fixed upstream wavelengths, others one or more tunable transmitters.

2.2. **MPCP.** EPON upstream medium access control is based on the multi-point control protocol (MPCP) standardized by IEEE (802.3ah). We consider the WDM extension to MPCP proposed by McGarry et al. [17].

| | | | |
|---|---|---|---|
| Destination address | 6 | Destination address | 6 |
| Source address | 6 | Source address | 6 |
| Length/Type = 88-08 | 2 | Length/Type = 88-08 | 2 |
| Opcode = 00-03 | 2 | Opcode = 00-02 | 2 |
| Timestamp | 4 | Timestamp | 4 |
| Number of queue sets | 1 | Number of grants/flags | 1 |
| Report bitmap | 1 | Grant 1 start time | 0/4 |
| Queue 0 report | 0/2 | Grant 1 length | 0/2 |
| Queue 1 report | 0/2 | ... | |
| ... | 0/2 | Grant 4 start time | 0/4 |
| Queue 7 report | 0/2 | Grant 4 length | 0/2 |
| | | Sync time | 0/2 |
| repeat for more | | Grant 1 wavelength | 0/1 |
| queue sets | | ... | |
| | | Grant 4 wavelength | 0/1 |
| Pad/Reserved | 0/39 | Pad/Reserved | 9/39 |
| FCS | 4 | FCS | 4 |

FIGURE 2. Format of REPORT and GATE messages with WDM extension from [17]

In addition to registration and configuration, the protocol fulfills the following three essential functions through the exchange of so-called GATE and REPORT packets (see Figure 2):

— the exchange of timing information allowing clock synchronization and precise measurement of the propagation time for each ONU;
— reporting to the OLT the current contents of ONU queues;
— granting transmission opportunities to the ONUs based on these reports.

The way REPORT and GATE messages are used to realize DBA is implementation dependent. As discussed in the introduction, we distinguish offline and online scheduling and REPORT-driven and GATE-driven online scheduling. Our proposal is to implement GATE-driven online scheduling, as described in Section 3 below.

2.3. **QoS.** An important function of DBA is traffic management and the fulfillment of service level agreements (SLAs) expressing QoS requirements. Experience shows that it is, to say the least, very difficult to efficiently satisfy a range of diverse SLAs in any multiservice network. On the other hand, the high capacity of optical access networks makes it rather easy to ensure excellent quality for all, except in situations of overload. In this context, we believe the following simple WDM PON traffic management strategy is adequate.

The OLT ensures ONUs receive a "fair" bandwidth share, by regularly issuing GATE messages to each ONU. The issuing rate might vary from one ONU to another to perform inter-ONU differentiation if required. Each GATE grants upstream transmissions depending only on the ONU's own reports, implementing an upper limit that might be ONU dependent. The ONU is left to manage its own intra-ONU service differentiation, using the grants as appropriate to meet latency and throughput requirements of traffic classes, end users, application flows or particular services. The DBA ensures cycle times between successive grants is small enough to satisfy the most stringent latency requirements. We do not further consider intra-ONU scheduling in this paper.

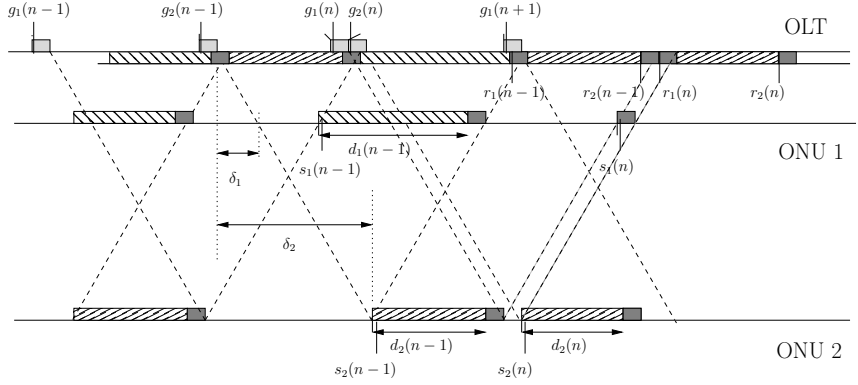

Figure 3. GATE-driven schedule with 2 ONUs

The GATE-driven DBA described next fulfills these objectives. We note, however, that GATE-driven scheduling would also be advantageous for more sophisticated QoS architectures.

## 3. GATE-DRIVEN SCHEDULING

We introduce the notion of GATE-driven scheduling in the context of the classical EPON with one wavelength in each transmission direction shared by a group of $N$ ONUs.

3.1. **Algorithm.** GATE-driven scheduling decouples GATEs from REPORTs, allowing the OLT to poll ONUs according to a freely defined periodic sequence. Let $\delta_i$ denote the one way propagation delay between the OLT and ONU $i$. For the sake of simplicity, we generally assume the schedule for emitting GATEs is round robin. The $n^{th}$ GATE sent to ONU $i$ has the nominal epoch $g_i(n)$ and instructs the ONU to send data for duration $d_i(n)$ starting at time $s_i(n)$. REPORT messages are piggybacked at the end of a granted data transmission starting, therefore, at tme $s_i(n) + d_i(n)$ or, if $d_i(n) = 0$, sent as an isolated packet at time $s_i(n)$. Let $r_i(n)$ denote the arrival time of the REPORT at the OLT. Duration $d_i(n)$ is determined from previously received REPORTs from ONU $i$. Each REPORT informs the OLT of data arrivals to the ONU queues since the time of the last REPORT while each GATE takes account of the latest information received (see Figure 3).

We apply a form of limited gated polling. Let the residual amount of time necessary to transmit the currently known backlog of ONU $i$ at time $t$ be $R_i(t)$. This is determined from previously received REPORTs and previously sent GATEs. The $n^{th}$ grant is upper bounded by $d_i^{max}$ such that

$$d_i(n) = \min\{R_i(g_i(n)), d_i^{max}\}.$$

Note that, if $d_i^{max} = \infty$, the ONU can send all traffic requested and we have a form of pure gated polling. In practice, GATE messages cannot always be sent at their designated time $g_i(n)$ since they share the downlink with data packets. Data are sent in encapsulated Ethernet frames. We assume the OLT schedules GATE messages with non-preemptive priority. The actual emission time can then differ from the desired epoch $g_i$ by up to the transmission time of a data packet. Assuming an MTU of 1500 bytes and a transmission rate of 1 Gb/s, the maximum delay is slightly more than 12 $\mu$s. Denote this time by $\tau$.

The time to send a REPORT message and the ensuing inter-packet guard time is $\Delta_R$. The corresponding time for GATE messages is $\Delta_G$. The guard time for upstream transmissions is around 1.5 $\mu$ yielding $\Delta_R \approx 2\mu s$ for a 1 Gb/s EPON. The downstream overhead $\Delta_G$ is smaller since all data is broadcast and guard times are not required. We later need the assumption $\Delta_G \leq \Delta_R$.

In the following, we suppose the ONU can always fully comply with the allocated grant $d_i$. In practice this may not be the case since Ethernet frames cannot be fragmented and the sum of the sizes of transmitted packets may need to be less than $d_i$. This will arise notably when $d_i = d_i^{max}$ or when the ONU sends recently arrived, higher priority packets that were not accounted for in REPORT messages. Such under-filling leads to a loss of traffic capacity that might be as high as 15% [10].

3.2. **Efficiency.** We seek to define the sequences $g$ and $s$ so that the upstream wavelength is always busy either transmitting data or REPORTs. Figure 3 illustrates a fragment of such a schedule for an EPON with two ONUs. The following proposition provides a recursive schedule that realizes this objective.

**Proposition 1.** *The following recursions define a feasible schedule and ensure the upstream link is fully utilized:*

$$g_i(n) = g_{i-1}(n) + d_{i-1}(n) + \Delta_R, \tag{1}$$
$$s_i(n) = g_i(n) + \Delta_G + \Delta_O - \delta_i. \tag{2}$$

*for $1 \leq i \leq N$, $n = 1, 2, \ldots$ where $\Delta_O$ is an offset satisfying $\Delta_O \geq 2\max\{\delta_i\} + \tau$, $g_0(1) = d_0(1) = 0$ and, by convention, quantities cycle as $g_0(n+1) = g_N(n)$.*

*Proof.* Feasibility requires

$$g_i(n) + \Delta_G + \tau \leq s_i(n) - \delta_i, \tag{3}$$

i.e., the GATE must arrive at the ONU before the scheduled start time accounting for maximal delay, and

$$g_i(n) \geq g_{i-1}(n) + \Delta_G, \tag{4}$$

i.e., the GATE for ONU $i$ cannot be emitted before the end of the preceding GATE. Condition (3) follows from (2) and $\Delta_O \geq 2\delta_i + \tau$ while (4) follows from (1), the assumption $\Delta_R \geq \Delta_G$ (see Sec. 3.1) and the fact $d_i(n) \geq 0$.

The uplink is always busy since

$$s_{i+1}(n) + \delta_{i+1} = s_i(n) + d_i(n) + \Delta_R + \delta_i, \tag{5}$$

the arrival time at the OLT of the first bit of the $n^{th}$ transmission from ONU $i+1$ (l.h.s) occurs immediately after the guard time following the arrival of the last bit of the $n^{th}$ transmission from ONU $i$ (r.h.s). This equation follows directly from (1) and (2). □

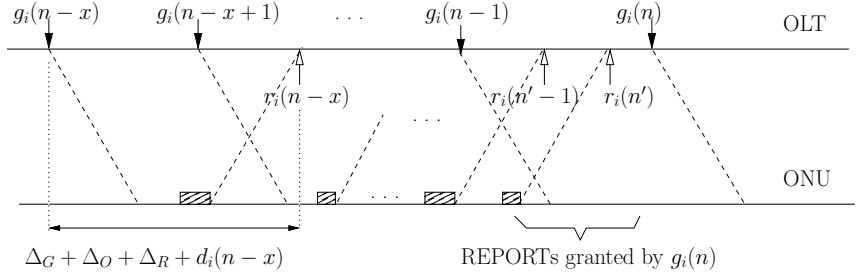

FIGURE 4. Correspondence of GATE and REPORT packets

3.3. **Traffic capacity.** The EPON under GATE-driven scheduling behaves like a single server polling system with fixed switchover time $\Delta_R$ with a non-standard service discipline. The grants do not relate to current queue occupancy but to the occupancy observed at some random time in the past as informed by the last REPORT to be received at the OLT. Figure 4 illustrates the interlacing of GATE epochs $g_i(n)$ with REPORT epochs $r_i(n')$ where, to simplify, we ignore durations $\Delta_G$ and $\Delta_R$. Note that the number of REPORTs arriving between $g_i(n-1)$ and $g_i(n)$ is random. It can take values between zero and an upper bound corresponding to the maximum cycle time $\sum(d_i^{max}+\Delta_R)$ divided by the minimum inter-REPORT gap $\Delta_R$. Known stability and cycle time results for periodic polling systems under a wide class of monotonic service disciplines carry over to this case [6].

**Proposition 2.** *The EPON with GATE-driven scheduling is stable iff*

$$(6) \qquad \rho + \max_i\{\rho_i/d_i^{max}\}S < 1,$$

*where $\rho_i$ is the ONU i load, $\rho = \sum_i \rho_i$ and $S = N\Delta_R$ is the sum of switchover times in one cycle. When the system is stable, the mean cycle time is*

$$(7) \qquad C = \frac{S}{1-\rho}.$$

*Proof.* It is intuitively clear that the random offset between GATEs and corresponding REPORTs in no way changes the stability conditions of the underlying polling system. Condition (6) thus follows directly from the corresponding result for periodic polling systems proved in [6]. Alternatively, the proposition may be seen as a simple corollary of Proposition 3 proved in Section 4 below.

The mean cycle time can be written

$$(8) \qquad C = \sum_i \mathrm{E}[d_i] + S.$$

Now, $d_i$ derives from a variable number of REPORTs received in previous cycles, as illustrated in Fig. 4. However, we do know there is one REPORT for every GATE and, when the system is stable, all reported data is eventually granted. We conclude that $\mathbb{E}[d_i]$ must be equal to the expected amount announced in an arbitrary REPORT. Applying Little's law, this is just $\rho_i C$. Formula (7) follows on making the substitution in (8). □

The fact that the grant sizes are limited is crucial to allow to other ONUs to remain stable even if some ONUs are unstable because of their load. This is called local stability in [6]. Let the ONUs be numbered such that $\rho_1 \leq \ldots \leq \rho_N$ and define $\hat{\rho}_j = \sum_{l=1}^{j} \rho_i$ as the overall load of ONUs $1, \ldots, j$. Let $S_j$ be the sum of the switchover time in one cycle and the deterministic maximum grants of ONUs $j+1, \ldots, N$ during one cycle. The sequence $(\hat{\rho}_j + \rho_j S_j/d_j)_j$ is increasing. Define

$$k = \max\{j, \hat{\rho}_j + \frac{\rho_j}{d_j} S_j < 1\}.$$

ONUs $k+1, \ldots, N$ are then unstable while ONUs $1, \ldots, k$ behave like a polling system with overall load $\hat{\rho}_k$ and switchover time during a cycle $S_k$.

Similarly, the formulas remain valid for non-homogeneous switchover times if $S$ is interpreted as the sum of switchover times in one cycle. This would arise if an ONU reported independent queue sets that are granted separately in one GATE message with therefore zero switchover times (up to four separate grants are possible: see Fig. 2). Finally, the periodic cycle does not need to be round robin. Some ONUs may be visited more frequently than others in each cycle, for example. We then need to calculate the total cycle switchover time $S$ and to interpret $d_i^{max}$ as the overall maximum service time for ONU $i$ in one cycle.

3.4. **Flexibility.** As noted above, GATE-driven scheduling allows the definition of arbitrary service orders and can therefore offer differentiated service rates to different ONUs. This might be useful when ONUs have different traffic levels or their owners/users have more or less costly subscriptions. Of course, increasing the visit frequency only increases throughput in case of saturation but it does reduce latency.

GATE-driven scheduling relies on repeatedly asking every ONU if it has packets to send, even if the ONU is persistently idle. Overhead might be reduced, thereby extending the capacity region, by reducing the polling frequency for ONUs that currently have no traffic. One possibility would be for the OLT to skip an ONU for $x$ cycles if it persistently reports no traffic. The number $x$ might grow (double, say) as the number of successively void reports increases. It would, of course, be necessary to maintain a minimal visit rate to avoid undue delay for an ONU that becomes active and to ensure latency remains acceptable for an active ONU transmitting at low rate. GATE-driven polling facilitates such flexible scheduling.

## 4. Capacity of a WDM PON

In this section we extend the GATE-driven performance results to a WDM EPON where ONUs have a full array of fixed wavelength transmitters. This solution is likely to be too expensive, however, and we therefore envisage an alternative where each ONU is equipped with a smaller number of fixed transmitters or a single tunable transmitter. The number of wavelengths is denoted by $L$.

4.1. **Full availability.** Each ONU can transmit without restriction on any wavelength. We distinguish two scheduling policies where the OLT implements either a separate but identical service cycle for each wavelength or a single cycle for all wavelengths. We assume the cycle in both cases is round robin.

4.1.1. *One cycle per wavelength.* The GATE epochs and transmission start times are determined for each wavelength using the recursions (1) and (2). It follows from Proposition 1 that these schedules are feasible and efficient. Traffic capacity is given by the following proposition.

**Proposition 3.** *The full availability WDM EPON with GATE-driven per-wavelength scheduling is stable iff*

$$\rho + \max_i \left(\rho_i/d_i^{max}\right) S < L, \tag{9}$$

*where $S = N\Delta_R$ is the total switchover time per cycle. When the system is stable, the mean cycle time for each wavelength is*

$$C = LS/(L - \rho). \tag{10}$$

*Proof.* Let $W_i(t)$ be the queue content of ONU $i$ at time $t$ (expressed as the time necessary to transmit this content at the wavelength rate) and define $X(t)$ to be the process $\{W_1(t), \ldots, W_N(t), Y(t)\}$ where vector $Y(t)$ collects the supplementary variables necessary to make $X(t)$ Markovian. Specifically, $Y(t)$ has the following components

— for each wavelength: the next ONU in the cycle to be sent a GATE message; the epoch, start time and grant of the last GATE message,
— for each ONU: the time since the last arrival; the epochs and values of the last $n$ reports where $n = \lceil \sum d_i^{max} + N\Delta_R)/\Delta_R \rceil$.

The latter quantity is chosen to ensure the process records all REPORTs that may still be relevant to compute the next grant.

The process $X(t)$ being Markovian, the fluid limit approach is used to demonstrate the sufficiency of (9). See [3], for example, and [7] in the case of multi-server polling systems. Fluid limits $\overline{W}_i(t)$ of the state of the queues at ONU $i$ for $1 \leq i \leq N$ is defined by

$$\overline{W}_i(t) = \lim_{|w| \to \infty} \frac{1}{|w|} W_i(|w|t),$$

where $w = (W_i(0))$. Since only the components $(W_i(t))$ of $(X(t))$ may can grow without bound, the Markov process will be shown to be ergodic if there exists some $T > 0$ such that $\overline{W}_i(T) = 0$ holds for all $i$, independently of its initial value $\overline{W}_i(0)$ (see Theorem 4.16 in [3]).

Let $G_i^l(t)$ be the total amount of transmission time granted to ONU $i$ by wavelength $l$ before $t$ with

$$G^l = \sum_i G_i^l, \quad G_i = \sum_l G_i^l, \text{ and } G = \sum_i G_i = \sum_l G^l.$$

Let $V_i^l(t)$ be the number of separate GATE messages sent to ONU $i$ before $t$ for wavelength $l$ with corresponding sums $V^l$, $V_i$ and $V$ (the wavelength "visits" the ONUs). Since all wavelengths realize the same round robin schedule, for the fluid limit we have, for $t > 0$ and $1 \leq l \leq L$,

$$\overline{V}_i^l(t) = \overline{V}^l(t)/N, \tag{11}$$

i.e., every ONU is visited with the same relative frequency by every wavelength.

Assume that at time $t$, some ONUs in the fluid system still have a backlog of work. Let $\mathcal{F}(t)$ be the subset of ONUs that still have a fluid backlog at $t$, $\mathcal{F}(t) = \{i : \overline{W}_i(t) > 0\}$.

Denoting time derivative with a dot, recalling that fluid limits are almost everywhere differentiable, we have

$$\dot{\overline{W}}(t) = \rho - \dot{\overline{G}}(t), \tag{12}$$

where $W(t) = \sum_i W_i(t)$. We now proceed to express $\dot{\overline{G}}(t)$ in two ways. First, since the wavelengths are fully utilized,

$$\dot{\overline{G}}(t) = L - \Delta_R \dot{\overline{V}}(t), \tag{13}$$

i.e., the useful transmission rate is the total capacity minus the switch overhead. Second, considering the rate at which data is transmitted,

$$\dot{\overline{G}}_i(t) = \rho_i, \text{ for } i \notin \mathcal{F}(t), \tag{14}$$

$$\dot{\overline{G}}_i^l(t) = d_i^{max} \dot{\overline{V}}_i^l(t), \text{ for } i \in \mathcal{F}(t). \tag{15}$$

Summing relations (14) and (15), we have:

$$\begin{aligned} \dot{\overline{G}}(t) &= \sum_{i \notin \mathcal{F}(t)} \rho_i + \sum_{i \in \mathcal{F}(t)} d_i^{max} \sum_l \dot{\overline{V}}_i^l(t) \\ &= \rho - \sum_{i \in \mathcal{F}(t)} \rho_i + \sum_{i \in \mathcal{F}(t)} d_i^{max} \dot{\overline{V}}(t)/N. \end{aligned} \tag{16}$$

Equating the r.h.s of (16) with that of (13), we have an equation for $\dot{\overline{V}}(t)$. Solving and substituting in (13) and (12) yields

$$\dot{\overline{W}}(t) = \frac{\sum_{i \in \mathcal{F}}((\rho - L)d_i^{max} + S\rho_i)}{\sum_{i \in \mathcal{F}} d_i^{max} + S}. \tag{17}$$

Now, due to condition (9), any term of the sum in the numerator of the r.h.s. of (17) is negative. This implies in particular that there exists some $T$ such that $\overline{W}(T) = 0$ independently of the initial value of $\overline{W}(0)$. The Markov process $(X(t))$ is therefore ergodic.

To prove the condition is necessary, we proceed as in [7]. Note that when the system is stable, by symmetry, the mean time between two consecutive visits of any wavelength to a given ONU is the same. This is the expected cycle time, $C$. Let $D_i^l$ be the expected overall transmission time granted to ONU $i$ by wavelength $l$ in a cycle that, for definiteness, starts and ends with the visit of wavelength 1 to ONU 1. By symmetry, $D_i^l$ is the same for each $l$, $D_i^l = D_i$, say. Thus $C = N\Delta_R + \sum_i D_i$. Equating the expected amount of work arriving in a cycle with the expected total grant, we have,

$$LD_i = \rho_i \left( S + \sum_{i=1}^N D_i \right).$$

Summing over $i$ gives an equation for $\sum_i D_i$, eventually yielding $D_i = \rho_i S/(L-\rho)$. Since, for $1 \leq i \leq N$, by construction $D_i < d_i^{max}$ we deduce the stated necessary condition for stability. Relation (10) is now straightforward to verify. □

4.1.2. *One overall OLT schedule.* Assume now that the OLT maintains a single round robin cycle and instructs the currently considered ONU, ONU $i$, say, to transmit on the next available wavelength taking account of grants issued in previous GATE messages. Let $\lambda$ be that wavelength and designate the ONU in question by $j_\lambda$. With a slight abuse of notation let $g_\lambda$, $s_\lambda$, $d_\lambda$ and $f_\lambda$ denote the corresponding GATE epoch, start time, grant and finish time, respectively. The finish time includes the piggybacked REPORT message and guard time: $f_\lambda = s_\lambda + d_\lambda + \Delta_R$. The following proposition defines a feasible and efficient schedule.

**Proposition 4.** *Let $g_i$ and $s_i$ be the GATE epoch and start time of ONU $i$, the next to be handled in the round robin cycle. The following relations define a feasible schedule and ensure all wavelengths are fully utilized:*

$$g_i = g_\lambda + d_\lambda + \Delta_R, \tag{18}$$

$$s_i = g_i + \Delta_G + \Delta_O - \delta_i, \tag{19}$$

*where*

$$\Delta_O = 2\max_i\{\delta_i\} + \tau.$$

*Proof.* It is straightforward to verify that (18) and (19) ensure the feasibility condition

$$g_i + \Delta_G + \tau \leq s_i - \delta_i, \tag{20}$$

i.e., the GATE arrives at the ONU before the scheduled start time even accounting for the maximum possible downstream delay $\tau$. In particular, the GATE can always be sent before $g_i + \tau$ on the downstream wavelength used for ONU $j_\lambda$.

To show the wavelengths are fully utilized we need to prove $s_i + \delta_i = f_\lambda$, i.e., the transmission from ONU $i$ arrives at the OLT immediately after the previous transmission. By (19) applied to ONU $j_\lambda$, we have $s_\lambda = g_\lambda + \Delta_G + \Delta_O - \delta_{j_\lambda}$. We also have $f_\lambda = s_\lambda + d_\lambda + \delta_{j_\lambda}$. Substituting for $s_\lambda$ and recognizing from (18) and (19) that $s_i = g_\lambda + d_\lambda + \Delta_G + \Delta_R + \Delta_O - \delta_i$, we readily verify that $f_\lambda = s_i + \delta_i$. □

**Proposition 5.** *The full availability WDM EPON with GATE-driven next available wavelength scheduling is stable iff*

$$\rho + \max_i (\rho_i/d_i^{max}) S < L, \tag{21}$$

*where $S = N\Delta_R$ is the total switchover time per cycle. When the system is stable, the mean cycle time is*

$$C = S^2/(L - \rho). \tag{22}$$

We omit the proof that closely mirrors that of Proposition 3. The key is again that each ONU receives the same number of visits. Instead of (11) we have $\dot{\overline{V}}_i(t) = \dot{\overline{V}}(t)/N$ for $1 \leq i \leq N$.

4.1.3. *Generalizations.* As for single wavelength EPON it is possible to extend the above capacity results. Conditions for local stability when some ONUs are unstable can be inferred by appropriately redefining switchover times and residual overall load. Non-homogeneous switchover times can be taken into account. The service cycle need not be round robin and some ONUs can be visited more frequently than others. However, crucially, when the OLT operates a separate cycle for each wavelength, the cycles should have the same visit frequency for each ONU to enable a relation equivalent to (11). The cycle can adaptively omit inactive ONUs to economize switchover ties and to alleviate the processing load of the ONU. Note that this flexibility is facilitated by GATE-driven scheduling.

4.2. **Limited availability.** While the ultimate aim is to evaluate the traffic capacity of a hybrid EPON where ONUs have various transmission capabilities, we so far have variable partial results when wavelengths are not fully available.

4.2.1. *Partial sharing.* Consider first the case where each ONU is equipped with a number of fixed wavelength transmitters that may be less than $L$. To simplify we first ignore the grant limits $d_i^{max}$ so that the underlying service discipline is gated polling. In this case the necessary and sufficient conditions for stability do not depend on the switchover times and are given by the following proposition.

**Proposition 6.** *When the service discipline is unlimited-gated, the WDM EPON with partial sharing is stable iff, for all subsets $\mathcal{O}$ of ONUs, we have*

$$\sum_{i \in \mathcal{O}} \rho_i < |\mathcal{L}(\mathcal{O})|, \tag{23}$$

*where $\mathcal{L}(\mathcal{O})$ is the set of wavelengths that can serve at least one member of $\mathcal{O}$.*

*Proof.* Necessity of (23) is obvious since at least one ONU queue would otherwise grow unboundedly. To prove sufficiency, we apply the fluid limit approach using the notation of Proposition 3. Recall that $\mathcal{F}(t)$ denotes the set of ONUs that still have a fluid backlog at $t$. Let $\mathcal{F}^l(t)$ be the subset of $\mathcal{F}(t)$ that can use wavelength $l$. Considering that the system is work conserving, we necessarily have,

$$\dot{\overline{G}}^l(t) = 1, \text{ for } l \text{ such that } \mathcal{F}^l(t) \neq \phi, \tag{24}$$

where $\phi$ denotes the empty set. Moreover, for $l$ such that $\mathcal{F}^l(t) \neq \phi$, the wavelength is necessarily busy serving some ONU from $\mathcal{F}(t)$. It follows that only wavelengths $l$ such that $\mathcal{F}^l(t) = \phi$ can be serving ONUs not in $\mathcal{F}(t)$. We conclude,

$$\sum_{l, \mathcal{F}^l(t)=\phi} \dot{\overline{G}}^l(t) = \sum_{i \notin \mathcal{F}(t)} \dot{\overline{G}}_i(t), = \sum_{i \notin \mathcal{F}(t)} \rho_i,$$

where we make use of (14). Thus

$$\dot{\overline{G}}(t) = \sum_{l, \mathcal{F}^l(t)=\phi} \dot{\overline{G}}^l(t) + \sum_{l, \mathcal{F}^l(t) \neq \phi} \dot{\overline{G}}^l(t), = \sum_{i \notin \mathcal{F}(t)} \rho_i + |\mathcal{L}(\mathcal{F}(t))|. \tag{25}$$

By (12),

$$\dot{\overline{W}}(t) = \sum_{i \in \mathcal{F}(t)} \rho_i - |\mathcal{L}(\mathcal{F}(t))|. \tag{26}$$

and since $\mathcal{F}(t)$ can be any subset of ONUs, we conclude $\dot{\overline{W}}(t)$ is negative under conditions (23) and the system is therefore stable. □

To see the potential impact of limited grants we consider a limiting case where switchover times are negligible and the grant limit tends to zero. In this case, wavelengths are used in head of line processor sharing mode – each ONU with a non-empty queue gains a certain share of the wavelengths it can use. We show by a simple example that the natural stability conditions are then not sufficient.

EXAMPLE 1.
Consider the case of two wavelengths $l_1$ and $l_2$ and two ONUs, such that ONU 1 can transmit on $l_1$ and $l_2$ but ONU 2 can transmit only on $l_2$. A natural condition for stability would be

$$\rho_2 < 1 \text{ and } \rho_1 + \rho_2 < 2. \tag{27}$$

We now proceed to show this condition is not sufficient. Assume condition (27) holds. If (27) holds, then either $\rho_1 < 3/2$ or $\rho_2 < 1/2$. If $\rho_1 < 3/2$, queue 1 empties with probability 1. As above, skipping technical details, let $(\overline{W}_1(t), \overline{W}_2(t))$ be a fluid limit of ONU queue contents. If $(\overline{W}_1(0), \overline{W}_2(0)) = (0, x)$ with $x > 0$, at the fluid scale $W_1$ is a birth and death process and the probability the queue of ONU 1 is empty is $\pi_0 = 1 - 2\rho_1/3$. The fluid equation for $(\overline{W}_2(t))$ is therefore given by

$$\dot{\overline{W}}_2(t) = \rho_2 - \frac{1}{2}(1 - \pi_0) - \pi_0 = \frac{1}{3}\rho_1 + \rho_2 - 1.$$

The necessary and sufficient condition for stability is thus (27) supplemented by $\rho_1 + 3\rho_2 < 3$.

This simple example shows that a loss of capacity occurs at queue 1: if $l_1$ were used efficiently, $l_2$ would have only to cope with load $\max(0, \rho_1 - 1)$ at queue 1, so that the stability condition would be $\max(0, \rho_1 - 1) + \rho_2 < 1$ which is precisely condition (27). It may be verified that this would be the case if, instead of equally sharing $l_1$, ONU 2 were given priority over ONU 1. This suggests there is scope for optimized scheduling but only when the grant size is not large enough that Proposition 6 is not a good approximation.

4.2.2. *Tunable transmitters.* As with partial sharing, it appears difficult to derive analytical formulas for traffic capacity of a WDM EPON where ONUs are equipped with tunable transmitters. The condition (9) is clearly necessary and would provide a reasonable capacity approximation when the probability an ONU can use multiple wavelengths in heavy traffic is small. Additional necessary conditions are obtained on considering each ONU in isolation. We clearly must have for $1 \leq i \leq N$,

$$(28) \qquad \rho_i < t_i \frac{d_i^{max}}{d_i^{max} + \Delta_R},$$

where $t_i$ is the number of tunable receivers in ONU $i$.

To investigate whether these natural conditions might also be sufficient, we consider another toy example. We again neglect switchover times and suppose the maximum grant size becomes very small.

EXAMPLE 2.
Consider the network with two wavelengths $l_1$ and $l_2$ and three ONUs. Each ONU can only use one wavelength at a time. The natural stability conditions now read

$$(29) \qquad \max(\rho_1, \rho_2, \rho_3) < 1 \text{ and } \rho_1 + \rho_2 + \rho_3 < 2.$$

The model is a multi-server multi-queue processor sharing system like that considered in [4].

Assume $\rho_1 > \rho_2 > \rho_3$ and (29) holds. This implies $\rho_3 < 2/3$ and the queue of ONU 3 empties with probability 1. At equilibrium, when the other ONUs are saturated, the probability that this queue is non-empty is given by $1 - \pi_0 = 3\rho_3/2$. The queue of ONU 2 then receives service at a rate no less than

$$\frac{2}{3}(1 - \pi_0) + \pi_0 = 1 - \frac{1}{2}\rho_3.$$

Consequently, queue 2 empties with probability 1 under the condition $\rho_2 < 1 - \rho_3/2$ which also holds as a consequence of (29). We conclude that, if (29) holds, then the process $(W_2(t), W_3(t))$ returns to state $(0, 0)$ with probability 1. Hence when ONU 1 is saturated, this process is ergodic. Let $p$ be the probability that both queues are non-empty under the invariant distribution. The fluid limit equation for

$(W_1(t))$ is then

(30) $$\dot{\overline{W}}_1(t) = \delta \stackrel{\text{def.}}{=} \rho_1 - \frac{2}{3}p - (1-p) = \rho_1 - 1 + \frac{1}{3}p.$$

Theorem 3.1 of [4] states that this quantity is negative under the additional condition $\rho_1 + \rho_2 + \rho_3 < 1$. This result can be improved as follows.

Assume $\rho_2 + \rho_3 < 1$. By comparing the process $(W_2(t), W_3(t))$ with the content of a simple queue with input rate $\rho_2 + \rho_3$ and service rate 1, it is straightforward to show that $W_2(t) + W_3(t)$ is stochastically smaller than the workload of the queue and consequently that $p$ is less than the probability this queue is non-empty, i.e., $p < \rho_2 + \rho_3$. In view of (30), $\delta$ is negative if $\rho_1 - 1 + (\rho_2 + \rho_3)/3 < 0$. We deduce the following sufficient stability condition

$$3\rho_1 + \rho_2 + \rho_3 < 3, \quad \rho_2 + \rho_3 < 1$$

which is weaker than the condition $\rho_1 + \rho_2 + \rho_3 < 1$ of [4]. It remains to establish if this condition is also necessary or, indeed, if necessary condition (29) is also sufficient.

Establishing the traffic capacity of the WDM EPON with tunable transmitters thus appears as a challenging largely open problem.

## 5. Performance

We evaluate the performance of GATE-driven and REPORT-driven scheduling in a long reach EPON. We use the stability limits and the mean cycle time derived in Section 4 and 5 for GATED-driven scheduling and simulation to obtain the mean packet delay for GATE-driven and the performance of REPORT-driven scheduling.

5.1. **Simulation set up.** We consider an EPON with a transmission bit rate of 1 Gbps per wavelength in both upstream and downstream directions. The total number of active ONUs is $N = 20$ and these generate fixed size 1000 byte packets according to a Poisson process. Each ONU has infinite buffer capacity. The size of REPORT and GATE messages is 64 bytes and the time guard is 1.5 $\mu$s (i.e $\Delta_G = \Delta_R \approx 2.12$ $\mu s$). One-way propagation delays $\delta_i, i = 1, \ldots, N$, are uniformly distributed between 10 and 500 $\mu$s corresponding to distances of between 2-100 km.

5.2. **GATE-driven vs REPORT-driven.** We first present results of a classical TDM EPON with a single wavelength in each direction.
Unlimited gated service. We compare GATE-driven and REPORT-driven scheduling without any limit on the grant size ($d\neg max_i$ is very large). Figure 5 shows the mean cycle time as a function of total load for symmetric traffic, i.e., all ONUs contribute equally to the total traffic load. As expected, both schedulers realize the full traffic capacity and the cycle time only becomes significant for very high loads ($> .95$). The cycle time of REPORT-driven scheduling is never smaller than 1ms, the longest round trip propagation tile of any ONU. This propagation time is experienced as a minimal delay for GATE-driven scheduling, as shown in Figure 6. This delay, excluding the transmission time, occurs even at low load due to the two stage reporting. Delays for REPORT-driven scheduling are 50% higher than for GATE-driven.

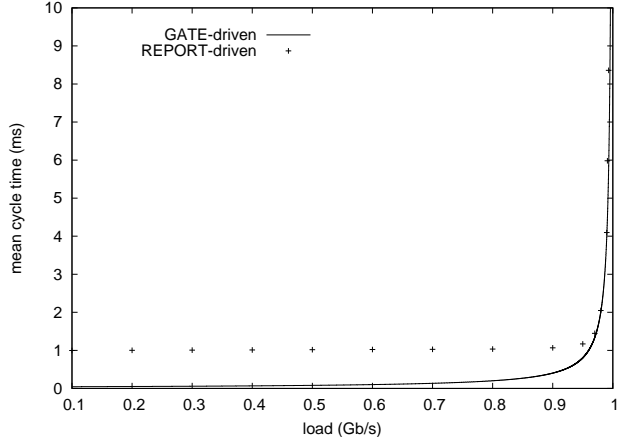

FIGURE 5. Mean cycle time with unlimited gated service (one wavelength).

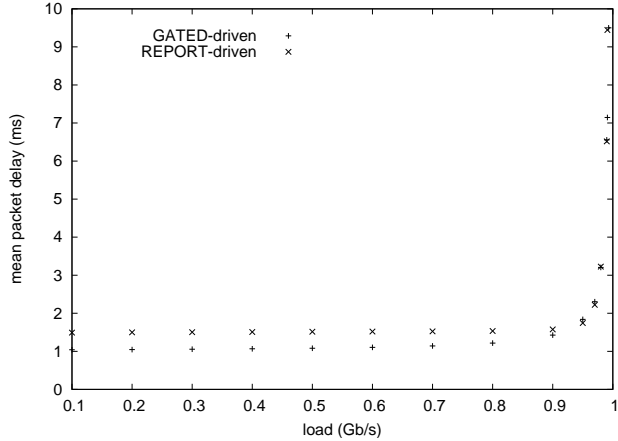

FIGURE 6. Mean packet delay with unlimited gated service (one wavelength).

Limited gated service. Figures 7 and 8 depict the mean packet delay of GATE- and REPORT-driven scheduling, respectively, for limited gated grant sizing with $d_i^{max}$ corresponding to 2, 4 and 6 Kbytes. The vertical lines in Figure 7 plot the stability limits given by equation (6). We see that the more efficient utilization of the upstream channel of the GATE-driven scheduling increases the stability limit substantially, especially for small maximum grant sizes. We also observe that GATE-driven scheduling presetrves the 50% delay advantage at low and medium loads. Mean cycle times are compared in Figure 9. The cycle time for REPORT-driven is again determined mainly by the propagation time with a slight increase as load tens to the capacity. The maximum cycle time for GATE-driven is equal to $N(\Delta_R + d^{max})$.

5.3. **Full availability WDM scheduling.** We present results the full availability WDM PON using a single round robin cycle per OLT. Results with one cycle per wavelength are very similar. Figure 10 represents the mean packet delay for GATE-driven scheduling for 2 and 3 wavelengths with a maximum grant size of 10Kbytes. We consider symmetric and asymmetric traffic loads. In the asymmetric traffic case, we set the arrival rate of 15 ONUs to $\gamma$ and that of the 5 other ONUs $5\gamma$ and vary $\gamma$ to cover the load range. The figure plots the mean delay of the most heavily loaded ONUs. The vertical lines are the stability limits for both traffic scenarios obtained through equation (6). The mean packet delay remains around 1 ms up to loads very close to capacity confirming that this stability limit is indeed the essential performance parameter for these systems. Figure 11 gives the mean cycle time for symmetric traffic depicting the maximum attained when every ONU transmits a maximal grant per cycle.

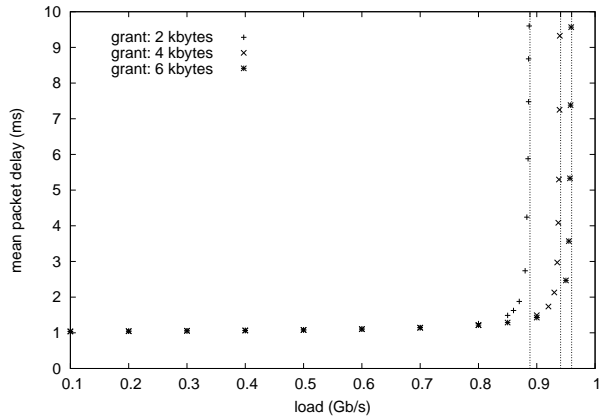

Figure 7. Mean packet delay with limited gated service for GATE-driven (one wavelength)

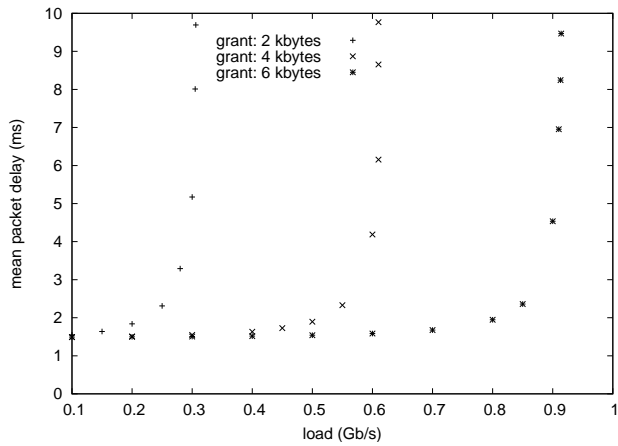

Figure 8. Mean packet delay with limited gated service for RREPORT-driven (one wavelength)

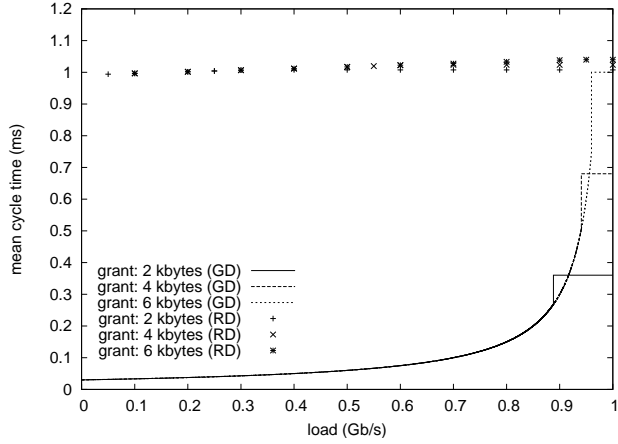

Figure 9. Mean cycle time with limited gated service (one wavelength)

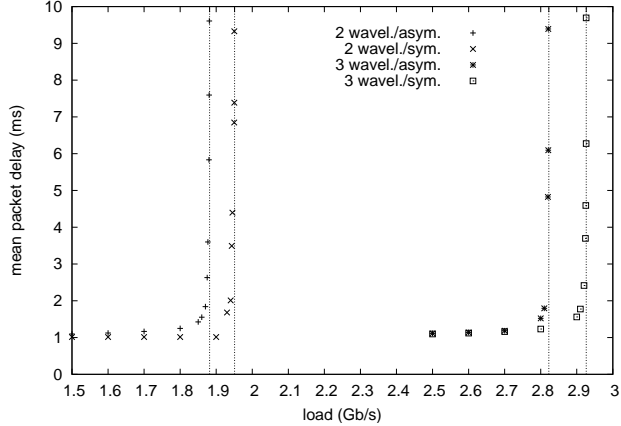

Figure 10. Mean packet delay with limited grant sizing for GATE-driven (full wavelengths availability)

5.4. **Limited availability WDM scheduling.** Figure 12 shows results for a limited availability system where ONUs are equipped with a tunable transmitter. We consider 3 wavelengths and set the grant size gated grant sizing to 8 kbytes. The number of active ONUs is now 4 and 20 and we consider both symmetric and asymmetric traffic. The asymmetry is such that 75% of ONUs have load $\gamma$ and 25% have load 5 $\gamma$. Vertical lines on the figure correspond to the capacity estimated using formula (21) giving the capacity of a corresponding full availability system. For the 20 ONU network, these analytical results are a good approximation mainly because in heavy load each ONU rarely has the opportunity to use more than one wavelength. For only 4 ONUs with traffic imbalance, formula (21) fails to approximate capacity. In fact, for this case, capacity is determined by the additional necessary condition $\rho_i < d_i^{max}/(d_i^{max} + \Delta_R)$ applied to the single heavily loaded ONU giving a capacity of 1.58 Gb/s.

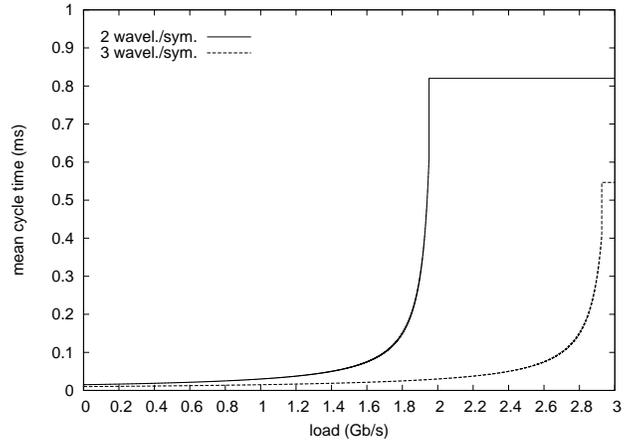

Figure 11. Mean cycle time with limited grant sizing for GATE-driven (full wavelength availability)

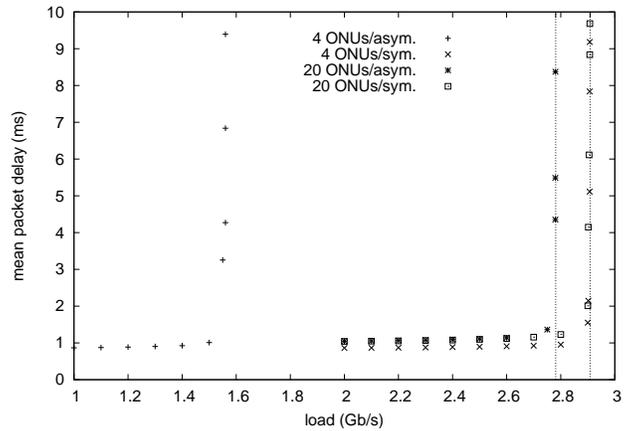

Figure 12. Mean packet delay with limited grant sizing for GATE-driven (limited availability)

## 6. Conclusions

The proposed GATE-driven DBA algorithm for long-reach WDM EPONs has been shown to have attractive properties in terms of both performance and flexibility. Traffic capacity is maximized while the mean packet delay is lower than that obtained with previously proposed online and offline schedulers. The algorithm is simple to implement and conforms to existing EPON standards. The algorithm allows flexible periodic service cycles that could be adapted dynamically depending on observed ONU activity. We have not yet explored the full potential of this flexibility.

We have derived closed-form traffic capacity formulas for a WDM EPON where all wavelengths are fully available to the ONUs. These formulas can constitute good approximations for more practically interesting networks where ONUs have limited access to wavelengths or can only use one wavelength at a time.

There is clearly considerable scope for further work. We would like to explore how one can exploit the potential for flexible scheduling to realize effective inter-ONU differentiation. It would be useful also to perform evaluations that confirm our belief that simple online scheduling is adequate to meet QoS requirements. In other words, we intend to show that per flow latency and throughput requirements can best be realized by intra-ONU differentiation that is independent of the DBA. It remains, of course, to more completely analyze the hybrid WDM EPON to derive useful dimensioning methods that properly account for the different capabilities of legacy and evolved ONUs that are controlled by the OLT.

(N. Antunes) INRIA Paris–Rocquencourt, France on leave of absence of University of the Algarve, Faro, Portugal
  *E-mail address*: `NAntunes@ualg.pt`

(Ch. Fricker, Ph. Robert, J. Roberts) INRIA Paris — Rocquencourt, Domaine de Voluceau, 78153 Le Chesnay, France.
  *E-mail address*: `Christine.Fricker@inria.fr`

  *E-mail address*: `Philippe.Robert@inria.fr`
  *URL*: `http://www-rocq.inria.fr/~robert`

  *E-mail address*: `James.Roberts@inria.fr`